\documentclass[11pt]{amsart}
\usepackage{geometry}                
\usepackage{graphicx}
\usepackage{amssymb}
\usepackage{epstopdf}
\usepackage{soul}
\usepackage{xcolor}

\newcommand{\be}{\begin{equation}}
\newcommand{\ee}{\end{equation}}
\newcommand{\fight}{3.0in}

\title[Measuring entropy of exchange economies]{Towards macroeconomic analysis without microfoundations: Measuring the entropy of simulated exchange economies\footnote{Corresponding author: R.S.MacKay@warwick.ac.uk}}
\author{Y. Luo, R.S.MacKay, Nick Chater}
\address{Warwick Business School and Mathematics Institute, University of Warwick}
\email{Yihang.Luo@warwick.ac.uk, R.S.MacKay@warwick.ac.uk, Nick.Chater@wbs.ac.uk}
\date{\today}                   

\begin{document}

\begin{abstract}

The theory of thermal macroeconomics (TM) analyses economic phenomena within the mathematical framework of classical thermodynamics, using a set of axioms that apply to the purely macroscopic aspects of an economy \cite{CM}. The theory shows that the possible macro-behaviours are governed by an entropy function.   In simple idealised cases, the entropy function can be calculated from the rules governing the interactions of individual agents. But where this is not possible, TM predicts that the entropy can nonetheless be measured empirically through an economic analogue of calorimetry in physics. We show using computer simulations the in-principle 
feasibility of this approach:~an entropy function can successfully be measured for a range of simulated economies that we tested.  In cases where entropy can be calculated analytically from microfoundational assumptions, the measured entropy agrees well.  In more complex cases, where microfoundational analysis is infeasible, our method of measuring entropy still applies and is validated by demonstrations that entropy is a state function of an economic system, i.e., exhibits path independence. This appears to hold even for some systems to which we don't have a proof that the Axioms of TM apply. Furthermore, in all cases tested, entropy is concave, as predicted by TM. As shown in \cite{CM}, once the entropy function is established for a simulated exchange economy, it is possible to derive prices, the value of money and various other quantities, and make predictions about the effects of putting two or more economies in contact.

\end{abstract}
\maketitle

\section{Introduction and motivation}

For more than half a century, one of the major trends in economics has been the effort to explain aggregate economic phenomena---such as prices, trade flows, and inflation---in terms of microfoundations, that is, the behaviour and interactions of individual economic agents such as people, households, and firms. However, the project of constructing such microfoundations is inherently challenging because each individual decision-maker is enormously complex. After all, each individual economic agent has roughly the complexity of the entire human brain, rather than being described as following simple rules; and behaviour of each agent is shaped by multiple layers of social norms, institutions, and interactions with the natural environment.

To make microfounded theories tractable, economists typically introduce radical simplifications, such as assuming perfect rationality and complete information. Yet this approach has fundamental limitations. First, the rational calculations posited in these models often exceed what is computationally plausible for human minds \cite{M24, S55}. Second, attempts to incorporate psychologically realistic decision mechanisms or bounded knowledge quickly become analytically intractable, even when the departures from full rationality are quite modest \cite{S55, A98}. 

One response to these difficulties has been to abandon the search for analytic solutions altogether and turn instead to agent-based simulations, which allow for more realistic modelling of heterogeneous and boundedly rational agents interacting in complex environments \cite{AF}.

In recent work \cite{CM}, we have developed an alternative approach, \emph{thermal macroeconomics} (TM), which maintains an analytical framework at the aggregate level, without specifying microfoundations. According to this approach, it should be possible to predict aggregate economic phenomena at the macro level, from a macro-quantity we call ``economic entropy'', even when the corresponding microfoundations are intractable to mathematical analysis and therefore lie beyond the scope of standard economic modelling. The parallel with the history of classical thermodynamics in the natural sciences in the nineteenth century suggests that this might be possible. Classical thermodynamics, which focuses entirely on macro-quantities such as pressure, volume, and temperature, was developed by Carnot, Joule, Clausius, Maxwell and Gibbs well before there was any common agreement even on the molecular theory of matter. Only with the advent of statistical mechanics (created by Maxwell, Boltzmann and Gibbs, among others), was it possible to construct microfoundations for particularly simple systems such as ideal gases, and to demonstrate a precise correspondence between the two levels of description. In particular, it was possible to show that in such simple systems the macroscopic concept of entropy could be identified with the logarithm of the `volume' of microscopic configurations (microstates) consistent with the macroscopic properties of the gas (such as its total energy, volume, and number of particles). Crucially for the present argument, classical thermodynamics also proved to be an extremely powerful tool for analysing physical, chemical, and biological systems---such as biochemical processes or climate dynamics---for which micro-founded models are typically intractable. Here, we ask whether the same pattern holds for economic systems.

TM maps certain key aggregate economic phenomena onto the mathematical structure of classical thermodynamics, building on the axiomatic foundation developed for classical thermodynamics by Lieb and Yngvason \cite{LY99}. As in physics, when applied to economics, this axiomatic framework operates purely at the macro level: specifying the aggregate behaviour of economic systems and their interactions without making microfoundational assumptions about the nature of individual agents. The key conclusion is that each economy has an ``entropy'' function of its macro-state, which governs the possible changes in state on putting several economies in contact. Namely, the total entropy can never decrease, and for any change of states that increases the total entropy there is a possible mode of contact (or sequence of such contacts) that achieves that change.

As an illustration, we studied the consequences of TM both analytically and through computer simulation, for a micro-founded model of individual agents that we call a Cobb–Douglas (CD) economy, in which agents engage in pairwise ``trades'' stochastically, governed by power-law utility functions \cite{CM, LMC}. The Cobb-Douglas model plays a role analogous to that of the ideal gas in the bridge between classical thermodynamics and statistical mechanics.  The paper \cite{LMC} extended the simulations to several variants of the CD economies, and corresponding theory for their TM entropy has been derived in \cite{M25b}. 

A key question, then, is whether TM can predict the behavior of an economic system for which the microfoundations are too complicated for mathematical analysis, and hence beyond the scope of the microfounded approaches typically used in economic theory. If so, this would imply that TM can provide an additional, and potentially powerful, method for understanding economic phenomena that are beyond the scope of current methods.

In this paper, we address this question in the context of a range of variants of Cobb–Douglas economies. For a Cobb–Douglas economy, the entropy function can be computed directly from the bottom up. But when the micro-foundations are not simple enough for their aggregate properties to be deduced, we may try instead to \emph{measure}, rather than calculate, the key phenomenological quantities such as entropy that govern the TM theory. The ability to measure quantities like the ``economic temperature'' and the ``value'' of a good in a variety of simulated economies was demonstrated in \cite{LMC}.  It remains, however, to demonstrate how to measure economic entropy.  A method to do this was sketched in \cite{CM}, analogous to calorimetry in physics and chemistry.  

The objective of the present paper is to test this approach in practice:~to demonstrate that the entropy function can be measured for simulated exchange economies, and to explore whether the system’s behaviour conforms to thermodynamic principles at the ``macro'' level, according to these measured quantities. The analogy with physics and chemistry is direct:~in many physical systems, such as chemical reactions \cite{A2023} or the data compiled in ``steam tables'' \cite{W2002}, entropy cannot be computed from first principles but is measured empirically. The second law of thermodynamics then allows the theorist to make very precise predictions about ``macro'' behaviour from these measurements, such as are used in the design of steam turbines. Similarly, in chemical engineering, entropy determines both whether a process can occur and the direction in which a process can proceed.

In this paper, we explore the economic context by using agent-based simulations to create a variety of ``model'' economic systems to be described at the macro-level---including cases where the agent-based systems are too complex to be analysed from the micro-level. 

We show that entropy, as well as related quantities, can indeed be measured at the macro level for simulated economies.
This constitutes the first crucial step in demonstrating that a thermal macroeconomic analysis can provide an autonomous level of explanation, independent of detailed microfoundations.

\section{Mathematical set-up}
\label{sec:setup}

Suppose a simulated exchange economy $E$ satisfies the axioms of \cite{CM}.  Then the key result of \cite{CM} is that it has an ``entropy" function of state that governs the allowed transitions on connecting it to other economies satisfying the axioms.
To keep things simple but not oversimplified, we suppose that there are at most three types of good, of which one is called money, and there are no internal barriers to movement of any of these.  Denote by $M$ the amount of money in the economy, and by $G, H$, the amounts of goods of types $g$ and $h$ (sometimes we will write $(G,H)$ as $(G_1, G_2)$ and use ``goods'' to refer to just them, not money).  We wish to measure its entropy function $S(M,G,H)$.  We can do this by an economic analogue of calorimetry.

We take our simulated economies to be of the forms used in \cite{LMC} or as extended in \cite{M25b} (or further extensions to be described here), where each agent has a ``utility function'' $u$ of their amounts of possessions (including perhaps their previous possessions or the possessions of other agents), and agents carry out pairwise exchanges with outcome probability density proportional to the product of their utilities, conserving their combined amounts.  Under fairly mild conditions in \cite{M25b}, 
the simplest of these economies satisfy the axioms for \cite{CM} and so have an entropy function.  Under these conditions, the entropy function is shown in \cite{M25b} to be the logarithm of an integral over all agents called a ``partition function'', but except in simple cases, an explicit expression for the resulting entropy can not be expected. 

An example where the entropy can be computed is a (homogeneous, 2-good) Cobb-Douglas (CD) economy.  It consists of a number $N$ of agents who have ``exponents'' $\eta,\alpha_g, \alpha_h > 0$ and a ``utility function'' 
\be u(m,g,h) = m^{\eta-1} g^{\alpha_g-1} h^{\alpha_h -1}
\ee
for their amounts $m$ of money and $g,h$ of the two types of good.  Pairs of agents $i,j$ encounter each other  independently at  rates $k_{ij}$, with the encounter rate graph being connected (that is, for all pairs of agents $i,j$, there is a path between agents $i=i_0,\ldots,i_l = j$ of some length $l$ with $k_{i_n i_{n+1}}>0$ for each $n$).
On encounter of $i$ with $j$, they pool their possessions and redistribute them between themselves with a probability density proportional to the product of their utilities for the outcome, subject to the constraints of non-negativity and conservation of the total amounts of each type, independently of everything else.  Such an economy (in the thermodynamic limit, where $N \to \infty$ with fixed mean amounts of the three types of good per agent) is simple enough that a formula can be derived for the entropy (extrapolating the analysis in \cite{CM} from two to three types of good):
\be S(M,G,H) = N\left(\eta \log\frac{M}{N} + \alpha_g \log\frac{G}{N} + \alpha_h \log\frac{H}{N}\right). \ee
The entropy is interpreted as an aggregate value of the economy.
By differentiation, from this are deduced (marginal) ``values'' for money and each type of good: 
\be \beta = \frac{\eta N}{M},\ \nu_g = \frac{\alpha_g N}{G},\ \nu_h = \frac{\alpha_h N}{H}. \ee  
The reciprocal $T = \frac{1}{\beta}$ is called ``economic temperature'' and the ratios $\mu_g = \frac{\nu_g}{\beta}, \mu_h = \frac{\nu_h}{\beta}$ are interpreted as ``market prices''.

For more general exchange economies, however, the computation of entropy might be too complicated, or we might not even have a proof that the axioms of \cite{CM} apply.  Instead one could hope to measure the entropy.
The key formula from \cite{CM} that allows us to measure entropy in an arbitrary exchange economy satisfying the axioms is
\be dS = \beta dM + \nu_g dG + \nu_h dH, \ee
for quasistatic changes in $M, G, H$.  Here, $\beta$ is the value of money in the economy and $\nu_j$ is the value of the good of type $j$.

We can measure $\beta$ and the $\nu_j$ by attaching to $E$ a CD economy and allowing exchange of money and goods between them (at equilibrium, the values for money become equal in the two economies, as do the values and hence the prices $\mu_j = \nu_j/\beta$ of each type of good).  This was already demonstrated in \cite{LMC} for a variety of micro-economic models. 

We call the attached CD economy a ``meter''. We adopt the method of \cite{LMC} whereby the money or goods entering or leaving $E$ is replaced immediately from an external reserve, so the measurement has no effect on $E$.  After equilibration between $E$ and the meter, we read off $\beta$ and the $\nu_j$ ($j=g,h$) in the meter economy as time-averages of $\eta/\bar{m}$ and $\alpha_g/\bar{g}, \alpha_h/\bar{h}$, respectively, where $\bar{m}$, $\bar{g}, \bar{h}$ are the mean money and goods of the two types per agent in the meter economy.

Our primary goal in this paper is to measure the entropy function $S(M,G, H)$ of $E$.
To measure it, we choose an arbitrary reference point $(M^*, G^*, H^*)$ and declare $S(M^*, G^*, H^*)=0$ (or some other value $S_0$ if desired). We measure its $\beta$ and $\nu_j$ using the CD meter.  Then we take a small step $(dM,dG, dH)$ to a point $(M',G',H')$, measure $\beta'$ and $\nu'_j$ there, and approximate $S(M',G',H')$ by 
\be S(M',G',H') \approx S(M^*,G^*,H^*) + \tfrac12 (\beta+\beta')dM + \tfrac12 (\nu_g + \nu_g') dG + \tfrac12(\nu_h+\nu_h')dH . \ee
By repeating this procedure, we can measure $S$ at a grid of points $(M,G, H)$.  One could just use one-sided approximation to the integration over a step but the above averaging of $\beta$ and $\nu_j$ over the two ends of the step increases the order of accuracy and requires virtually no more work over a grid, because the measured values $\beta', \nu_j'$ can be reused at the next step.

A crucial feature of the TM theory that we test is that the value of $S(M,G,H)$ should not depend on the route by which $(M,G,H)$ is reached from $(M^*,G^*,H^*)$.  Because of measurement error this will not be exactly true, so to make a good estimate of the function $S$, we compute the values of $(\beta,\nu_j)$ at all grid points and then make a least-squares fit to a function $S$ such that
\be S(M',G',H')-S(M,G,H) = \tfrac12(\beta+\beta') dM + \tfrac12(\nu_g+\nu_g') dG + \tfrac12(\nu_h+\nu_h')dH \ee for all edges in the grid.  

The failure to satisfy path-independence can be quantified by the least squares error, that we call residual sum of squares (RSS).\footnote{The least squares error can be considered in the infinitesimal grid size limit as the norm-squared of the divergence-free component of the Helmholtz-Hodge decomposition (HHD)  of the vector field $(\beta,\nu_g,\nu_h)$ on the space of $(M,G,H)$ with inner product to make $dM, dG, dH$ have equal norm (the HHD writes an arbitrary vector field $v$ as the sum of a gradient field $\nabla S$ and a divergence-free one $w$).} The appropriate quantity with which to compare the least squares error is the sum of squares of the changes before fitting an entropy function, that we call total sum of squares (TSS). 
If the ratio RSS/TSS is small then the fit is good (one can be more sophisticated with statistical significance tests taking into account the sample size, which is the number of edges).  So we call this ratio the ``goodness of fit''.

One prediction of \cite{CM} that we test is that the entropy function is concave. It would be good to test other predictions of the TM theory for the types of economy presented in this paper, as we did for some in \cite{LMC}, but we leave that for later work.

For systems for which we do not have a proof that they satisfy the conditions for the TM theory, we can nonetheless attempt to measure the entropy in the same way.  The goodness of fit (quantifying path-independence of the measured entropy) is then a strong indicator of whether the theory is likely to apply.

To complete the presentation of the set-up, we recall that
as argued in \cite{CM}, we expect money in real economies to be ``pure,'' meaning that for fixed amounts of goods, the value of an amount of money is proportional to the fraction of the total amount of money it represents, and prices are proportional to the total amount of money. 
It follows that the entropy function for an economy with pure money has the form \begin{equation}
S=N(\eta \log \bar{m} + R(\bar{g}_1,\bar{g}_2))
\label{eq:pure}
\end{equation}
for some $\eta>0$ and function $R$ \cite{CM}. A counter-intuitive consequence is that the value of money is independent of the amounts of goods in the economy, even if the only use for money is to facilitate exchange of goods.
We chose several of our simulated economies to have this property, which is easily imposed by taking
agents to have utility function of the form $u(m,g_1,g_2) = m^{\eta-1} U(g_1,g_2)$ for some function $U$.  In this case, we need only to test the dependence of the entropy on $G_1$ and $G_2$.  One could check that indeed they do have the form (\ref{eq:pure}) of entropy function, but it seemed to us that checking it was a point of minor significance.

\section{Verifying the computational method}
\label{sec:verify}

Before proceeding to the main tests of interest, we first check that the computational method recovers the known result for a few models.

For each case in this section, the test has two steps:~(i) we fit an entropy function by the procedure of the previous section, reporting on the goodness of fit (representing the degree to which path independence holds); (ii) we compare the fitted function to the true one, bearing in mind that there is an arbitrariness of additive constant for entropy of a system (the latter is easily dealt with by shifting one or other entropy function to make them have the same mean over the chosen grid).  

The second step comes with its own quantification that we call ``goodness of agreement'', namely the sum of squares of differences of entropy between the fitted function and the true one (once the above shift has been performed). The goodness of agreement can be quantified by the sum of squares of differences, compared to the total sum of squares of deviations of the true entropy from its mean.  Again, one can be more sophisticated, taking into account sample size (number of grid points in this case) and quoting a $p$-value, but the results are sufficiently clear that we do not pursue this here. 

\subsection{CD economies}
We begin by verifying that the method recovers the known entropy for two CD economies.
This is tautological if one uses the CD economy as its own meter, because then the relation 
\be dS = \frac{\eta}{\bar{m}} dM + \frac{\alpha_1}{\bar{g}_1} dG_1 + \frac{\alpha_2}{\bar{g}_2} dG_2, \ee 
with $N$ agents, gives 
\be S = N(\eta \log \bar{m} + \alpha_1 \log \bar{g}_1 + \alpha_2 \log \bar{g}_2), \ee 
up to an additive constant, which is the known result \cite{CM}.  But one can check it using a different CD meter. 

We tested a homogeneous CD economy and an inhomogeneous one.  The latter  means that different agents may have different exponents.  In both cases we used a CD meter with exponents $\eta=2 , \alpha_1=2 , \alpha_2 =2 $. 

\begin{figure}[htbp] 
   \centering
\includegraphics[height=\fight]{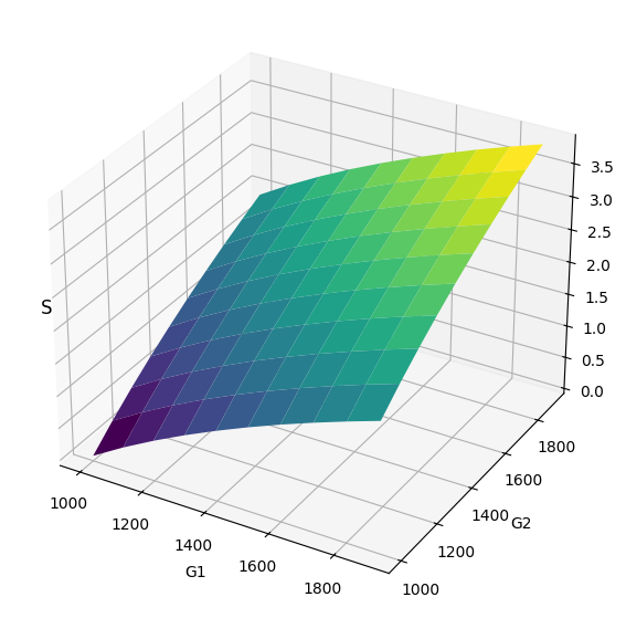} 
   \caption{Measured entropy per agent for a homogeneous CD economy with  $\eta=3 , \alpha_1=3 , \alpha_2 =3 $, as a function of amounts of the two types of good (with $M=1000$). 
   The goodness of fit is $8.34 \times 10^{-6}$.  The goodness of agreement is $9.87 \times 10^{-7}$.
   }
   \label{fig:CD1}
\end{figure}

For the homogeneous CD economy, we took exponents $\eta=3 , \alpha_1 =3 , \alpha_2 =3$.   The resulting measured entropy function is shown in Figure~\ref{fig:CD1}.  We see that the measurement procedure works to high accuracy.

\begin{figure}[htbp] 
   \centering
\includegraphics[height=\fight]{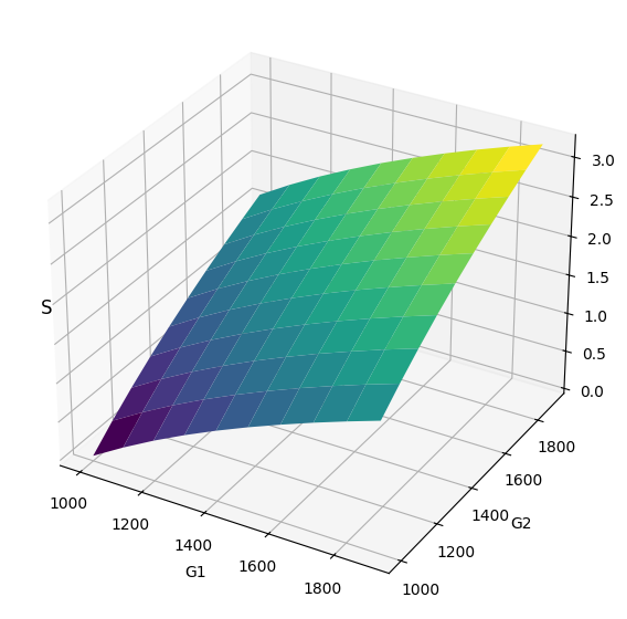} 
   \caption{Measured entropy per agent for a heterogeneous CD economy with  half of the agents with $\eta=3 , \alpha_1=3 , \alpha_2 =3 $  and half of the agents with $\eta=2 , \alpha_1=2 , \alpha_2 =2 $, as a function of amounts of the two types of good (with $M=1000$). The goodness of fit is $9.18 \times 10^{-6}$.  The goodness of agreement is $9.15 \times 10^{-6}$.}
   \label{fig:CD2}
\end{figure}

Next we took a heterogeneous CD economy in which a fraction  50\% of the agents have exponents  $\eta=3 , \alpha_1=3 , \alpha_2 =3 $ and the remainder have exponents $\eta=2 , \alpha_1=2 , \alpha_2 =2 $. The measured entropy function is shown in Figure~\ref{fig:CD2}.
The known result for the heterogeneous economy is
\be S(M,G_1,G_2) = \sum_i \eta^i \log\bar{m} + \sum_i \alpha_1^i \log\bar{g}_1 + \sum_i \alpha_2^i \log \bar{g}_2, \ee 
where the sums and means are over agents $i$ in the tested economy (derivation of this formula is a simple extension of what was done in an Appendix to \cite{CM}, or one can apply the more general derivation of \cite{M25b}). So again we find excellent agreement.


\subsection{Substitutes and complements economies}
Next we tested two cases where the entropy function is known, but in the form of a Legendre transform, 
\begin{equation}S(M,G,H) = \min_{\beta,\nu_g,\nu_h} \beta M + \nu_g G + \nu_h H - F(\beta,\nu_g,\nu_h),
\label{eq:Leg}
\end{equation}
for a ``free energy function'' $F$, rather than explicitly (the name comes from Physics, but is used in various senses there, often as only a partial Legendre transform). 
This is almost as good as the entropy being an explicit function, however, especially as $\beta, \nu_g, \nu_h$ in the above are precisely the ``values'' for $M,G,H$, which are measured quantities.  
It includes cases of simple models of substitutes agents and complements agents.

As introduced in \cite{LMC},
{\em substitutes agents} are ones whose utility depends on the amounts of goods of types 1 and 2 through only their sum, e.g.~
\be U(m,g_1,g_2) = m^{\eta-1}(g_1+g_2)^{\alpha-1} \ee 
for some $\alpha>0$; thus the two types are equivalent for the agent (think of two brands of essentially the same good).  {\em Complements agents} are ones whose utility depends only on the minimum of $g_1$ and $g_2$, e.g.~
\be U(m,g_1,g_2) = m^{\eta-1}\min(g_1,g_2)^{\alpha-1}; \ee thus the two types of good are useful to the agent only when present in equal amounts, the remainder having no use (think of left and right gloves).  Intermediate variants could also be constructed.

For only substitutes agents or only complements agents (of the above simple forms), we computed the entropy function in \cite{LMC} (using results of \cite{M25b}), as the Legendre transform of  
\begin{align} F(\beta,\nu_1,\nu_2) &= N\left(\eta \log \beta + \log\frac{\nu_1-\nu_2}{\nu_2^{-\alpha}-\nu_1^{-\alpha}}\right) \mbox{ (substitutes) } \\
F(\beta,\nu_1,\nu_2) &= N\left(\eta \log\beta + (\alpha-1)\log(\nu_1+\nu_2) + \log \nu_1 + \log \nu_2\right) \mbox{ (complements) }
\end{align}
Actually, in the case of complements, one can do the Legendre transform explicitly \cite{M25b}, but the formulae are somewhat involved so we do not write them out fully. 

One can also compute the free energy function for a CD economy.  The result (up to an additive constant) is 
\begin{equation}
F(\beta,\nu_1,\nu_2) = N(\eta \log\beta + \alpha_1 \log \nu_1 + \alpha_2 \log \nu_2).
\label{eq:F}
\end{equation}
These three examples have the special feature that 
\be S(M,G,H) = NC - F(\beta,\nu_1,\nu_2) \ee 
for some constant $C$, which allows the slight shortcut of ignoring the $\beta M + \nu_g G + \nu_h H$ in (\ref{eq:Leg}).
So to test the fit to the free energy functions, we can use the measured values of $\beta, \nu_j$ at the grid points and compare the fitted $S(M,G,H)$ with the theoretical $-F(\beta, \nu_1, \nu_2)$, shifted to give them the same means over the grid.

To make more substantial verification tests than a substitutes or complements economy on its own, we took a mixture of substitutes, complements and CD agents.  Their free energies add up, so the total is easy to compute, but it would be really messy to write the formula for the resulting entropy function, so this makes it a good test.

\begin{figure}[htbp] 
   \centering
\includegraphics[height=\fight]{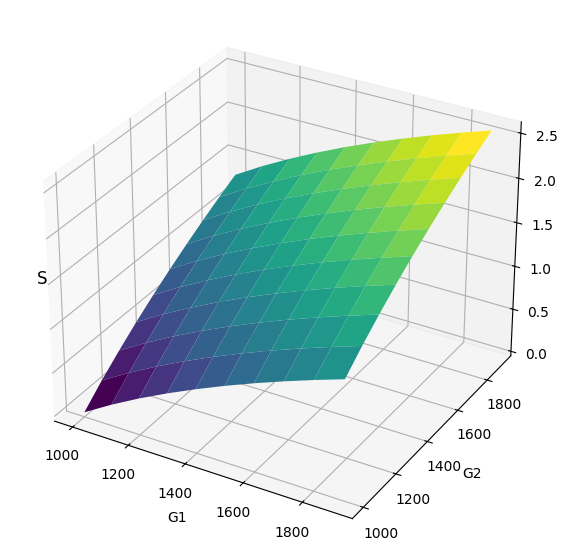} 
   \caption{Measured entropy per agent for an economy of 500 substitutes and 500 complements agents, as a function of amounts of the two types of good. The goodness of fit is $9.5 \times 10^{-6}$. The goodness of agreement with the theoretical free energy function is $4.64 \times 10^{-3}$.}
   \label{fig:SC}
\end{figure}

First, we took a mixture of substitutes and complements agents.  We took 500 of each and chose $\alpha=\eta=3$ for both types of agent.   
The resulting fit for entropy as a function of $G_1, G_2$ (with $M$ fixed at 1000) is shown in Figure~\ref{fig:SC}.
  
\begin{figure}[h!] 
   \centering
\includegraphics[height=\fight]{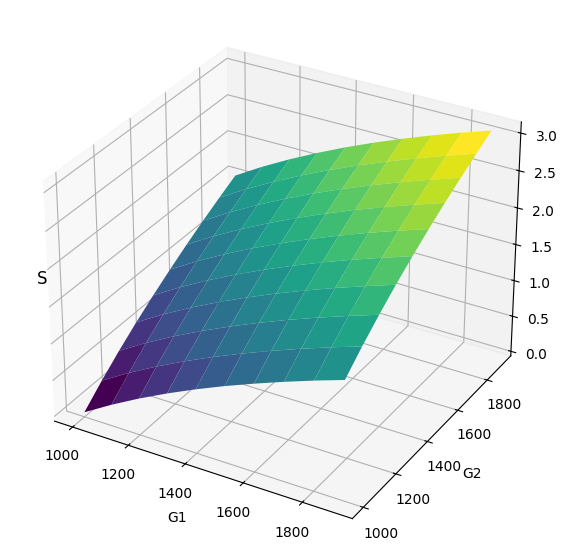} 
   \caption{Measured entropy per agent for an economy of 300 substitutes, 300 complements and 400 CD agents, as a function of amounts of the two types of good. The goodness of fit is $1.081 \times 10^{-5}$.  The goodness of agreement is $8.72 \times 10^{-4}$.}
   \label{fig:SCcd}
\end{figure}

Next, we changed to a mixture of 300 substitutes, 300 complements and 400 CD agents, giving the CD agents exponents $\alpha_1 = \alpha_2 = 3$.  The resulting measured entropy function is plotted in Figure~\ref{fig:SCcd}. 
In both these examples, we see excellent agreement with the predictions of the thermal economic account.

\subsection{Remarks}
In this section, we have verified that our computational method for measuring entropy works at the ``macro'' level:~for a selection of cases where we know the answer, the entropy calculations obtained are path-independent to high accuracy and agree with entropy calculations from microfoundations. 

\section{Tests for intractable simulated economies}
\label{sec:intractable}

We now apply the method to some cases for which the microfoundations are too complex for a bottom-up microfoundations analysis to be mathematically tractable, thus showing a proof of concept for the potential value of the TM approach as an autonomous level of description of economic systems.

We tested  
some economies in which agents have decreasing utility for too much of a good, or want to exchange only if the price is favourable, or have utility depending on the state of their neighbours in some graph. 

In each case, our test consists in just the first step from the previous section:\ fitting an entropy function and evaluating the goodness of fit (path independence, as quantified in Section~\ref{sec:setup}).

\subsection{Satiable agents}
\label{sec:satiable}
First, we tested an economy in which agents want to have at least a certain amount of the good of type 2 but are not interested in having a lot of it.  We call them {\em satiable} agents. 

\begin{figure}[htbp] 
   \centering
\includegraphics[height=1.5in]{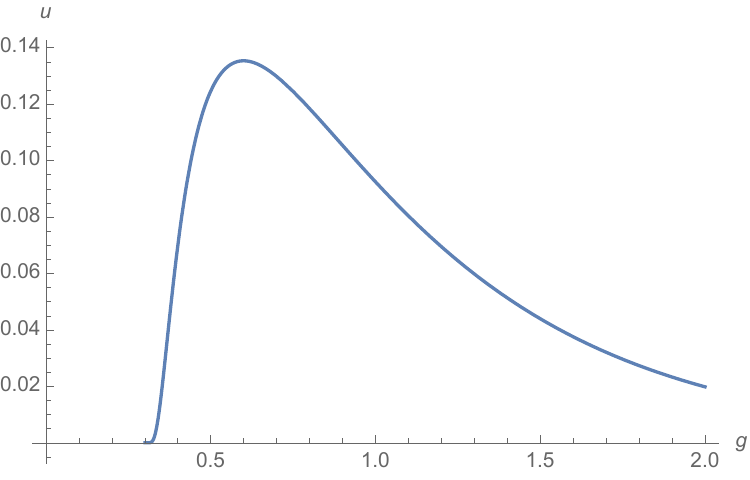} 
   \caption{Utility function $U_1$ for satiable agents, with $c=0.3, k=0.6$. }
   \label{fig:U2}
\end{figure}

We took $U(g_1,g_2) = g_2^{\alpha-1} U_1(g_1)$ with
\be U_1(g) = \exp\frac{-g^2}{k(g-c)}, \ee 
where $c = 0.3, k=0.6$, and $U_1(g)=0$ if $g\le c$.  This is plotted in Figure~\ref{fig:U2}. The partition function looks impossible to evaluate for this example, and even the free energy is given by an integral that Mathematica can't do.

\begin{figure}[h!] 
   \centering
\includegraphics[height=\fight]{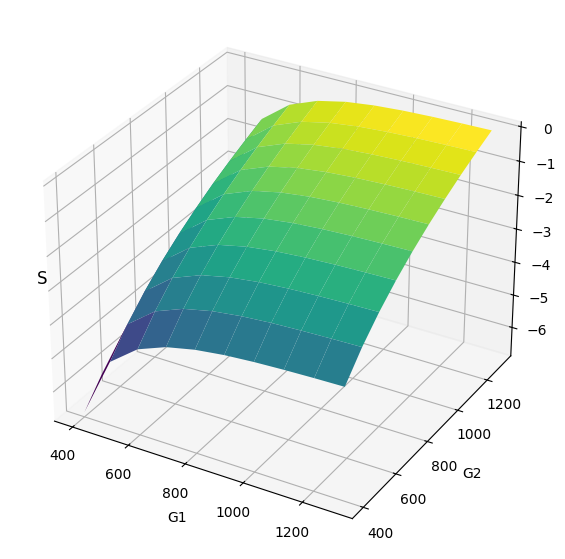} 
   \caption{Measured entropy per agent for an economy of 1000 satiable agents for good 1, as a function of amounts of the two types of good (with money $M=1000$).  The goodness of fit is $1.034 \times 10^{-5}$. 
}
   \label{fig:ex1}
\end{figure}

The measured entropy per agent is shown in Figure~\ref{fig:ex1} as a function of amounts of the two types of good. 
We see an excellent fit to an entropy function.
Note too that, as predicted by TM, the resulting entropy function is concave.

\subsection{Price-sensitive agents}
\label{sec:pricesens}
Next, we tested an economy in which agents prefer not to sell below some price nor to buy above some other price.
We call these {\em price-sensitive} agents.  
Note that money is not necessarily pure for such examples, because the dynamics depend on some price thresholds.

To set the scene, first consider the case where a trade is rejected (by one or other agent) if the implied price is not in the interval $(\mu_1,\mu_2)$, and the utility for the outcome is otherwise CD with  $\alpha=\eta=3$.  We consider just one type of good.
We denote the utility function for the outcome $(g',m')$ for an agent, conditional on their current $(g,m)$ by $u(g',m' | g,m)$.  Thus we take
\be u(g',m' | g,m) = g'^{\alpha-1} m'^{\eta-1} \mbox{ if } \mu_1 \le \frac{m'-m}{g-g'} \le \mu_2, \ee
and 0 otherwise.  
This is a case that does not fit into the framework of \cite{M25b}, because the probability distribution for the outcome of an encounter depends not only on the sum of the agents' possessions but also on the amounts individually. 

\begin{figure}[htbp] 
   \centering
\includegraphics[height=\fight]{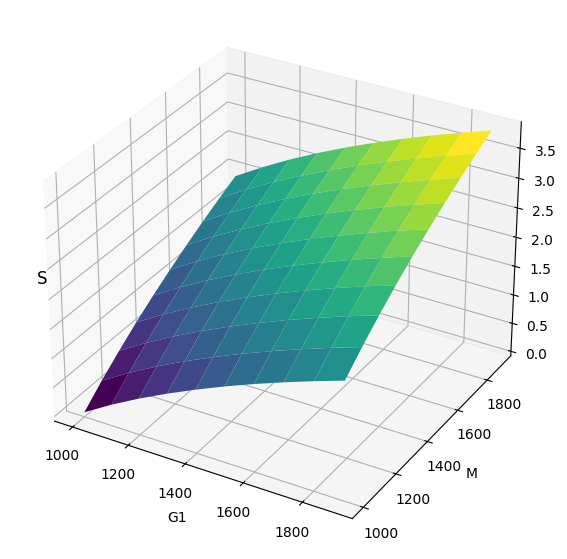} 
   \caption{Measured entropy per agent for an economy of 1000 agents who buy and sell only if the implied price is in $(0.9, 1.1)$, with a CD utility.  The goodness of fit is $7.84 \times 10^{-6}$.
}
   \label{fig:ex2}
\end{figure}

So we tested it, obtaining the measured entropy in Figure~\ref{fig:ex2}, with good path-independence. 
We suspect, however, that the entropy for this example is just
\be S = N(\eta \log M + \alpha \log G). \ee
The reason is that the resulting Markov process is reversible with respect to the CD stationary probability distribution, as can be checked quite easily.  Assuming the analysis of \cite{M25b} can be extended to this case, the entropy is then the CD entropy.  So we compared the measured entropy to the CD entropy and found a goodness of agreement $1.23 \times 10^{-5}$.

 
In particular, for this example, money turns out to be pure after all.
Modification from hard to soft thresholds determined purely by the implied price of the exchange does not affect the stationary distribution, by the same argument as above.

\begin{figure}[htbp] 
   \centering
\includegraphics[height=\fight]{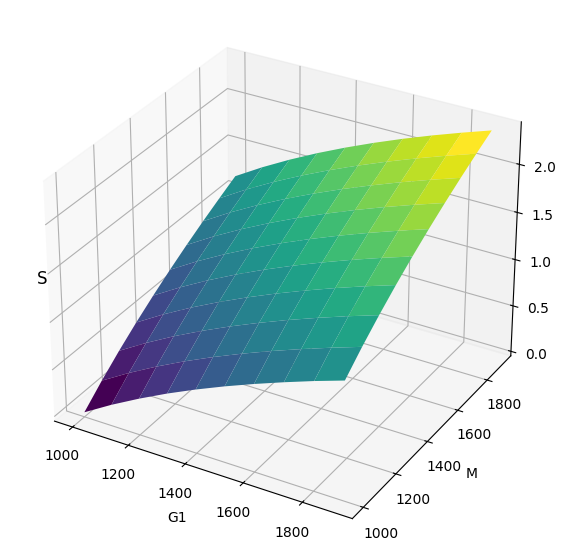} 
   \caption{Measured entropy per agent for an economy of 1000 agents who buy and sell only if the implied price is in $(0.9, 1.1)$, with the price-sensitive utility in Equation (\ref{eq:mod1}). 
   The goodness of fit is $1.567 \times 10^{-5}$.
}
   \label{fig:ex2a}
\end{figure}

To make more substantial tests, we modified the example in two directions.  The first modification is to make the distribution of volume of exchange depend on current amounts and implied price in a way that is not just a product of functions of outcome and implied price.  We think this  leads to an intractable stationary distribution.  So we tried 
\begin{equation}
u(g',m'|g,m) = (g+g')^{\alpha-1} (m+m')^{\eta-1} \mbox{ if } \mu_1 \le \frac{m'-m}{g-g'} \le \mu_2,
\label{eq:mod1}
\end{equation}
0 otherwise. The measured entropy is shown in Figure~\ref{fig:ex2a}.  Again agreement with the path-independence required for economic entropy to be a state function is good, and the resulting entropy is concave.

\begin{figure}[htbp] 
   \centering
\includegraphics[height=\fight]{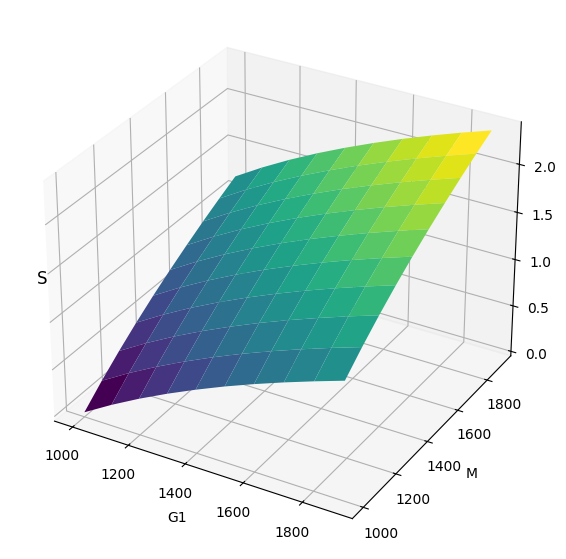} 
   \caption{Measured entropy per agent for an economy of 1000 agents with 200 of each of five types, who buy and sell only if the implied price is in $(0.5, 1.5)$, $(0.6,1.4)$, $(0.7,1.3)$, $(0.8,1.2)$ and $(0.9, 1.1)$, respectively, with the price-sensitive utility (\ref{eq:mod1}).
The goodness of fit is $1.128 \times 10^{-5}$.
}
   \label{fig:ex2b}
\end{figure}

A second modification is to suppose different agents have different price thresholds.  In such a case, it is possible for two agents to make an exchange that can not be reversed.  For example, suppose agent $i$ has thresholds $\mu^i_1 < \mu^i_2$ (for selling and buying, respectively\footnote{We do not allow thresholds in the other order, even though that might be considered rational, else the agent's possessions never decrease in the corresponding partial order.  We consider it more realistic for fortunes to rise and fall.}) 
and agent $j$ has thresholds $\mu^j_1 < \mu^j_2$ with $\mu^i_1 < \mu^j_2$ and $\mu^j_1 > \mu^i_2$. Then $i$ can sell to $j$ at any price in $(\mu^i_1, \mu^j_2)$ but $j$ can't sell to $i$ because $\mu^j_1 > \mu^i_2$. 
So the Markov process ceases to be reversible and there is no easy way to see what stationary distribution(s) it has.
See Appendix A for a discussion. There is a risk that there is no stationary distribution or that some agents end up with nothing or some agents end up with everything. But we tried a simulation and did not observe this arising in practice.  The measured entropy function is shown in Figure~\ref{fig:ex2b}. 
Again we see good evidence of path-independence.

One interesting property of price-sensitive agents is that they never give away money for nothing, so for example, putting two economies of price-sensitive agents in financial contact will not lead to any flow of money between them, even if at different temperatures.  This might suggest that they will fall outside the axiomatic framework of \cite{CM}. Nonetheless, our simulations suggest that an exchange economy with price sensitive agents might be well-described by the thermal macro-economic approach.

\subsection{Interdependent agents}
\label{sec:interdep}

The last case we tested was an economy in which each agent's utility depends on the possessions of their neighbours.  We call them {\em interdependent} agents.
This is a category of agents which in general we do not yet know to fit in the framework of \cite{CM}, so it makes a very interesting test. It is also interesting from a psychological and economic point of view, given that people's subjective valuation of their holdings of many types of good seems determined primarily by ``social comparison'' rather than than the absolute quantity of the good \cite{F05, V99}.

We took agents to be arranged in a circle, possessions to be money and one type of good, and 
\be
u(m,g,g_n) = m^{\eta-1}g^{\alpha-1} U(g-g_n),
\label{eq:interd}
\ee
where $g_n$ is the average amount of goods in a ``comparison'' neighbourhood, which we took to be  the $k=2$  nearest neighbours, and $U$ is some positive function.  
Each agent trades at the same encounter rate with a subset of agents that can be the whole set or its comparison neighbourhood or some other subset, as long as the encounter graph is connected.  It was simplest for the simulations to take the trading neighbourhood to be disjoint from the comparison neighbourhood, so we did that.  If not, then the probability distribution for the outcome has to include the fact that each agent's possessions occur in the utility for the other, which complicates the simulations (although it is otherwise unproblematic).   

\begin{figure}[htbp] 
   \centering
\includegraphics[height=\fight]{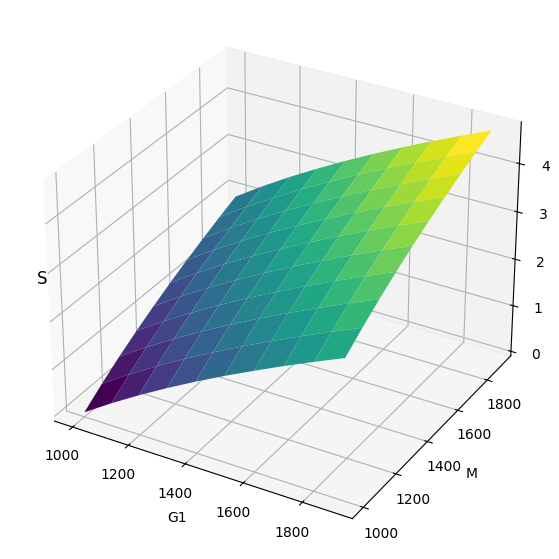} 
   \caption{Entropy measured for an economy in which agents' utilities depend on neighbours' possessions via (\ref{eq:interd}) with $U(x) = e^x$.  The goodness of fit is $7.31 \times 10^{-6}$. The goodness of agreement is $8.94\times 10^{-6}$
}
   \label{fig:ex3}
\end{figure}

We started with $U(x) = \exp x$.  The measured entropy function is shown in Figure~\ref{fig:ex3}. The fit to an entropy function is excellent again. 
But we later realised that the entropy can be computed exactly for this case.  The reason is that for an encounter between non-neighbours $i$ and $j$ the outcome density is proportional to $m_i^{\eta-1}g_i^{\alpha-1} e^{g_i} m_j^{\eta-1}g_j^{\alpha-1} e^{g_j}$, because the effect of possessions of neighbours is just multiplication by a constant factor $\exp (-g_{n_i}-g_{n_j})$, $n_i$ and $n_j$ denoting the neighbourhoods of $i$ and $j$ respectively.
Similarly, in interaction with an external agent (e.g.~in a CD meter), the factor $e^{-g_n}$ has no effect (but the factor $e^{g_i}$ does).
So if all encounters are between non-neighbours then the resulting system is the same as that for utility function \be \tilde{u}(m,g) = m^{\eta-1} g^{\alpha-1}e^g. \ee 
The free energy function for this can easily be computed to be \be F(\beta,\nu) = N (\eta \log \beta + \alpha \log(\nu-1))\ee plus an irrelevant constant.  So $\beta = N\eta/M$ and $\nu = 1+ N\alpha/G$, resulting in
\be S(M,G) = N(\eta\log \bar{m} + \bar{g} + \alpha \log \bar{g})
\ee
plus an irrelevant constant.
Indeed, we compared our measurements to $\nu = 1 + \alpha/\bar{g}$ and found goodness of agreement $1.06 \times 10^{-6}$.

\begin{figure}[htbp] 
   \centering
   \includegraphics[height=2in]{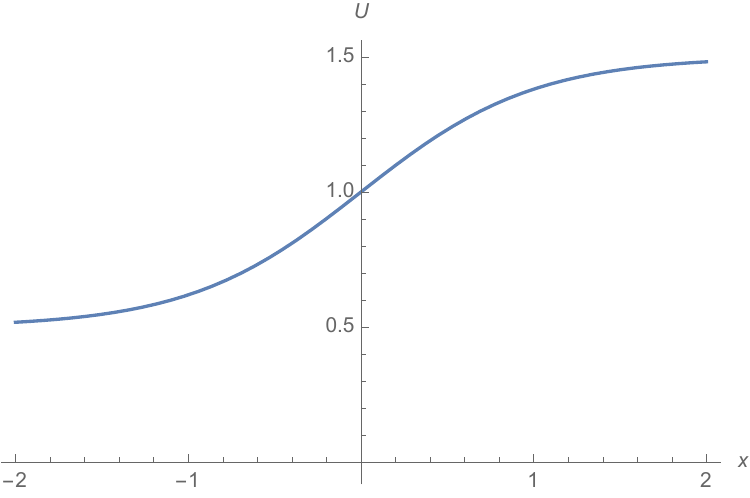} 
   \caption{Function $U$ with $a=3/2, b=1/2$.}
   \label{fig:plot}
\end{figure}

\begin{figure}[htbp] 
   \centering
\includegraphics[height=\fight]{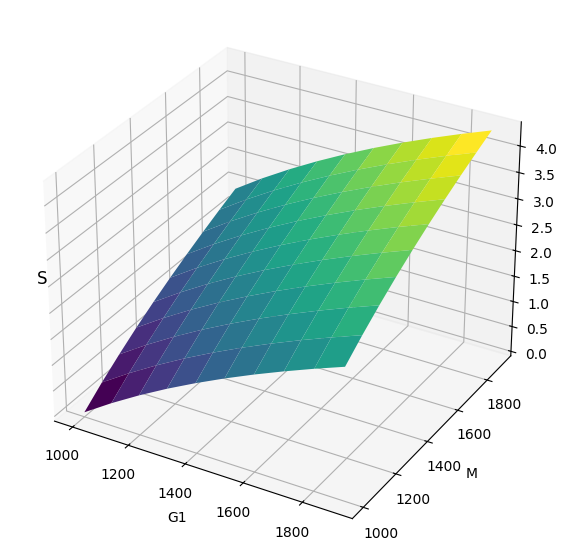} 
   \caption{Entropy measured for an economy in which agents' utilities depend on neighbours' possessions using (\ref{eq:tanh}).  The goodness of fit is $6.94 \times 10^{-6}$.
}
   \label{fig:ex3a}
\end{figure}

To create what we think is a case for which micro-level analysis is intractable, we switched to 
\be U(x) = \frac{a e^x + b e^{-x}}{e^x + e^{-x}}
\label{eq:tanh}
\ee
with some $a>b$ (plotted in Figure~\ref{fig:plot}), for which we do not know a stationary distribution. 
The resulting fit to an entropy function is shown in Figure~\ref{fig:ex3a}.
The path-independence is again excellent (and the resulting entropy concave), suggesting again that the scope of the TM theory is wider than the cases that we can justify so far.

\subsection{Remarks}
With these three types of example (plus those in Section~\ref{sec:verify} and in \cite{LMC}), we have only begun to scrape the surface of the range of possible models which may be amenable to the thermal macroeconomic approach: i.e., that we measure an entropy function, and the behaviour of the economy (i.e., prices, value of money, flows of money and goods and so on) is governed by the second law, just as in physics.  One could take mixtures of the types of agent, or invent other types of agent, or exit the realm of pairwise encounters, e.g.~consider exchange economies based on a central exchange with an order book (as is realistic for trading shares).  Going further, it would be interesting to see to what extent one could include models of agents as used in realistic economic simulations \cite{AF}.
But we hope that the examples treated here make a convincing case that entropy can be measured for many simulated economies.

\section{Discussion}
\label{sec:disc}
The paper has confirmed that the entropy function of simulated economies can be measured by an economic analogue of calorimetry, as was proposed in \cite{CM}.  A variety of types of exchange economy was tested. We found excellent agreement in cases where we know the answer. In all cases where we know the theory applies we confirmed that the required integral was indeed path-independent.  For cases where we do not know the theory applies, we could still apply the measurement method and we found excellent path-independence, strongly suggesting that the scope of the theory is much broader than we are currently able to prove.
As a bonus, in each case the entropy was found to be concave, as predicted by the theory. 

An interesting feature of our tests is that most of the measured entropy functions look very similar, up to scaling (an exception being for the satiable agents).  This may be partly an artefact of our choice of examples, but it is worth commenting that the entropy for any of these examples is independent of the encounter rate matrix (as long as it is connected).  
The relative insensitivity of the entropy function to changes in the micro-foundational factors may be highly significant for macroeconomics. First, it may suggest usefulness and robustness to detailed micro-level assumptions of the thermal economic approach; second it suggests that there will be autonomous macro-level generalisations which are not evident from considering micro-level assumptions, or, indeed, from applying any current methods in macroeconomics. The possibility that an entropy-based explanation may be valuable even when micro-foundational assumptions are both only partially known and highly complex, as when predicting real economies, is particularly promising.

It would be useful to carry out tests of the TM theory on the examples studied here, analogous to the tests made in \cite{LMC} on simpler examples. {This is a good direction for future work.} A further interesting question for future investigation is what happens if one attempts to measure the entropy for a system that does not satisfy the axioms, and so perhaps it does not have an entropy satisfying the conclusions of TM?  An example might be a ``Bouchaud'' economy \cite{LMC} in the parameter regime where ``condensation'' occurs, due to herding \cite{GBB}. We {hope to consider this issue} in a future paper.

In the longer term, a key question is how practical this method would be to implement on real economies. Firstly, the axioms of \cite{CM} will not all be satisfied exactly.  For example, real economies almost certainly violate \emph{extensivity} (Axiom A4 in \cite{CM}), because economies of different sizes will typically have some qualitatively distinct features. Moreover, given the scope of the current formulation of TM, one would need to restrict to exchange economies but perhaps none really exist, though the economy around a trading exchange is a good instance if one can consider it as a closed system. Nonetheless, we hope in future work to develop the theory to include production, consumption, finance, and so on, which might allow similar methods to those used here to be applied. 

To apply the economic analogue of calorimetry to measure the entropy of a real economy would require being to be able to make changes in a real economy, in particular injecting or removing money or stocks.  
Economics shares with other observational sciences the difficulty of controlled experimentation \cite{F53, H01}, and even basic structural quantities are typically inferred indirectly through econometric assumptions \cite{L83}. Simulation-based and agent-based approaches therefore provide a natural complementary methodology \cite{AF}, just as modelling plays a central role in observational sciences \cite{OSB}.

Finally, to apply these methods, one must often allow some time for the system to equilibrate. Some critics object that true equilibrium is never fully attained, e.g.~\cite{He}\footnote{This problem is arguably more severe for conventional economic theory, which analyses putative deterministic equilibria in economic systems, or in models such as Dynamic Stochastic General Equilibrium (DSGE) models, widely used in practical economic modelling by policy makers, which allow stochastic shocks, but still rely on the assumption that the economy remains in or near equilibrium at all times \cite{CET}.}, but this concern overlooks the importance of timescales. For the TM approach to be useful, it is sufficient that the time-scale for significant external perturbation to the system’s probability distribution is long compared with the time it takes for the system to relax close to equilibrium. In other words, as long as the perturbed system settles close to equilibrium before conditions shift again, the analysis remains valid. This is an application of the idea of local thermodynamic equilibrium.

\section*{Acknowledgements}

NC gratefully acknowledges support from the European Union/UKRI under Horizon Europe Programme Grant Agreement no.~101120763 – TANGO; ESRC UKRI/NSF - Grant Ref:~ES/Z504397/1; ESRC grant UKRI1676;  and Advanced Research and Invention Agency (ARIA) Collaboration Agreement SCNI-PR01-P16. Views and opinions expressed are however those of the author(s) only and do not necessarily reflect those of the European Union or the European Health and Digital Executive Agency (HaDEA). Neither the European Union nor the granting authority can be held responsible for them. For the purpose of open access, the author has applied a Creative Commons Attribution (CC-BY) licence to any Author Accepted Manuscript version arising from this submission.

\section*{Data availability}
The code for the simulations in this paper is available at \cite{Luo}.

\section*{Appendix A: Conditional utilities}
\label{app:condutility}

Our price-sensitive agents (Section~\ref{sec:pricesens}) introduce a case of conditional utilities, in which the utility for the outcome for an agent depends on its current state directly, not just through the sum of the amounts of goods of the agent and the agent it encounters.  We continue to use utilities in the sense of unnormalised densities for the probability distribution of the outcome of an encounter.  Conditional utilities make a mild extension of our class of Markov processes, 
but formally they are not covered in the treatment of \cite{M25b}.  Extending that to conditional utilities is a good project for the future.

Here, we make a beginning, around the question of the stationary probability for the process, which is necessary for the discussion about price-sensitive agents. 

Let $u_i(p_i'|p_i)$ be the conditional utility for agent $i$ for outcome $p_i'$ (vector of amounts of possessions) conditional on current possessions $p_i$.  If this does not grow too fast with $p'_i$, there is a vector $k$ such that 
\be \rho_i(p_i) = \int u_i(p'_i | p_i)e^{-k^T p'_i}\, dp'_i \ee converges for each agent $i$. Modify the utility of each agent $i$ to $u_i(p'_i|p_i) e^{-k^Tp'_i}$.
The factor $\exp(-k^Tp'_i)$ does not change the Markov process, because total possessions are conserved at each encounter, so the outcome probability density is unchanged.
We can then define  \be \sigma_i(p'_i,p_i) = u_i(p'_i|p_i)\rho_i(p_i). \ee  Suppose that in encounters \be \sigma_i(p_i',p_i) \sigma_j(p_j',p_j) = \sigma_i(p_i,p'_i) \sigma_j(p_j,p_j'). \ee  Then the Markov process is reversible with respect to the  distribution with density $\Pi_l\, \rho_l(p_l)$, restricted to the simplex of given total amounts $P$ of each type of possession.  Hence, this distribution is stationary. A factor $\exp(-k^TP)$ can be removed, to make up for the artificial inclusion of $e^{-k^Tp'_i}$.
Probably the 
entropy can then be obtained as the logarithm of the normalisation constant of the resulting distribution, as in \cite{M25b}.
But if this reversibility condition is not satisfied then there is no easy way to see a stationary distribution.
The reversibility condition is satisfied if each $\sigma_i$ is symmetric, but we suspect that the product formulation above gives a little more flexibility.

Extension 
to longer history-dependence than just the current state can be envisaged.

\end{document}